\documentclass[conference]{IEEEtran}
\IEEEoverridecommandlockouts
\usepackage{cite}
\usepackage{amsmath,amssymb,amsfonts}
\usepackage{algorithmic}
\usepackage{graphicx}
\usepackage{textcomp}
\usepackage{xcolor}
\def\BibTeX{{\rm B\kern-.05em{\sc i\kern-.025em b}\kern-.08em
    T\kern-.1667em\lower.7ex\hbox{E}\kern-.125emX}}

\usepackage{float}
\graphicspath{ {figures/} }

\usepackage{balance}

\usepackage{xspace}
\usepackage{booktabs} 
\usepackage{multirow}
\usepackage{caption}
\usepackage{multirow}
\usepackage{tabularx}
\usepackage{array}
\usepackage{amsmath} 
\usepackage{siunitx}
\usepackage{makecell}
    
\usepackage{titlesec}
\usepackage{chngcntr}

\newcounter{subsubsubsection}
\counterwithin{subsubsubsection}{subsubsection}
\renewcommand\thesubsubsubsection{\arabic{subsubsection}.\alph{subsubsubsection})}

\titleformat{\subsubsubsection}
  {\normalfont\normalsize\itshape}{\thesubsubsubsection}{1em}{}
\titlespacing*{\subsubsubsection}
  {0pt}{3.25ex plus 1ex minus .2ex}{1.5ex plus .2ex}

\newcommand{\subsubsubsection}[1]{%
  \refstepcounter{subsubsubsection}%
  \textit{\arabic{subsubsection}.\alph{subsubsubsection})\ #1}\hspace{0pt}%
}

\usepackage[pdftex,colorlinks=true,pdfstartview=FitV,
 linkcolor=black,citecolor=black,urlcolor=black]{hyperref}
\usepackage{needspace}

\usepackage{paralist}
\usepackage{ifthen}
\usepackage[normalem]{ulem} 
\usepackage{xcolor}

\newboolean{showedits}
\setboolean{showedits}{true} 
\ifthenelse{\boolean{showedits}}
{
	\newcommand{\del}[1]{\textcolor{red}{\sout{#1}}} 
	\newcommand{\nbe}[3]{
		{\colorbox{#3}{\bfseries\sffamily\scriptsize\textcolor{white}{#1}}}
		{\textcolor{#3}{\sf\small$\blacktriangleright$\emph{#2}$\blacktriangleleft$}}}
}{
	\newcommand{\del}[1]{} 
	
	\newcommand{\nbe}[3]{}
}
\usepackage[most]{tcolorbox}
\ifthenelse{\boolean{showedits}}
{
  \newtcolorbox{inserted}{%
       title=Inserted text:,
       colframe=blue,colback=blue!5!white,
       breakable,
       leftrule=0mm, 
       bottomrule=0mm,
       rightrule=0mm,
       toprule=0mm,
       arc=0mm, outer arc=0mm,
       oversize
  }
  \newtcolorbox{deleted}{%
       title=Deleted text:,
       colframe=red,colback=red!5!white,
       breakable,
       leftrule=0mm, 
       bottomrule=0mm,
       rightrule=0mm,
       toprule=0mm,
       arc=0mm, outer arc=0mm,
       oversize
  }
  \newtcolorbox{refactored}{%
       title=Rewritten text:,
       colframe=blue,colback=red!5!white,
       breakable,
       leftrule=0mm, 
       bottomrule=0mm,
       rightrule=0mm,
       toprule=0mm,
       arc=0mm, outer arc=0mm,
       oversize
  }
}{

}
\usepackage{amssymb}
\newboolean{showcomments}
\setboolean{showcomments}{true}
\newcommand{\id}[1]{$-$Id: scgPaper.tex 32478 2010-04-29 09:11:32Z oscar $-$}

\ifthenelse{\boolean{showcomments}}
{\newcommand{\nbc}[3]{
 {\colorbox{#3}{\bfseries\sffamily\scriptsize\textcolor{white}{#1}}}
 {\textcolor{#3}{\sf\small$\blacktriangleright$\emph{#2}$\blacktriangleleft$}}}
 }
{\newcommand{\nbc}[3]{}
 }

\newcommand{\ie}{\emph{i.e.},\xspace}

\newcommand{\etal}{\emph{et al.}\xspace}

\begin{document}


\title{\title{The Hidden Costs of Automation: An Empirical Study on GitHub Actions Workflow Maintenance} 
\thanks{}
}


\author{\IEEEauthorblockN{Pablo Valenzuela-Toledo\textsuperscript{1,4}, Alexandre Bergel\textsuperscript{2}, Timo Kehrer\textsuperscript{1}, Oscar Nierstrasz\textsuperscript{3}}
\IEEEauthorblockA{\textsuperscript{1}\emph{Software Engineering Group, University of Bern}, Bern, Switzerland\\
\textsuperscript{2}\emph{RelationalAI}, Bern, Switzerland\\
\textsuperscript{3}\emph{feenk GmbH}, Wabern, Switzerland\\
\textsuperscript{4}\emph{Universidad de La Frontera}, Temuco, Chile}
}

\maketitle


\begin{abstract}

GitHub Actions (GA) is an orchestration platform that streamlines the automatic execution of software engineering tasks such as building, testing, and deployment. Although GA workflows are the primary means for automation, according to our experience and observations, human intervention is necessary to correct defects, update dependencies, or refactor existing workflow files. In fact, previous research has shown that software artifacts similar to workflows, such as build files and bots, can introduce additional maintenance tasks in software projects. This suggests that workflow files, which are also used to automate repetitive tasks in professional software production, may generate extra workload for developers. However, the nature of such effort has not been well studied.
This paper presents a large-scale empirical investigation towards characterizing the maintenance of GA workflows by studying the evolution of workflow files in almost 200 mature GitHub projects across ten programming languages. 
Our findings largely confirm the results of previous studies on the maintenance of similar artifacts, while also revealing GA-specific insights such as bug fixing and CI/CD improvement being among the major drivers of GA maintenance.
A direct implication is that practitioners should be aware of proper resource planning and allocation for maintaining GA workflows, thus exposing the ``hidden costs of automation.''
Our findings also call for identifying and documenting best practices for such maintenance, and for enhanced tool features supporting dependency tracking and better error reporting of workflow specifications.

\end{abstract}

\begin{IEEEkeywords}
Continuous Integration,
GitHub Actions, 
Workflow Maintenance,
Empirical Study
\end{IEEEkeywords}

\section{Introduction}

The software industry has widely adopted Continuous Practices (CP) such as Continuous Integration and Delivery (CI/CD) to automate software engineering tasks~\cite{golzadeh2022rise}. The premise driving these practices suggests that they can help minimize integration issues, enable frequent integration, automatically deploy changes, and speed up feedback loops to software developers~\cite{fowler2006continuous,humble2010continuous,fitzgerald2017continuous}. 

Within the GitHub ecosystem, developers can automate software engineering tasks through GitHub Actions (GA), a widely adopted tool for implementing CP~\cite{golzadeh2022rise}. 
GA uses \texttt{YAML} workflow files as the building blocks for automating software engineering tasks. 
Developers can specify these files to perform jobs when a specific trigger event occurs throughout the definition of executable processes. The workflows can execute jobs concurrently, sequentially, or following a specific order defined by the developer. 
Each job comprises sequentially executed steps that embody commands, similar to bash scripts, to perform a common CI/CD task. 
Events that trigger workflows include pushing changes to a repository, generating new pull requests, and regularly scheduled events.

Although workflow files are the primary means for specifying automation, human intervention is necessary to correct defects, update dependencies, or refactor existing workflow files. For instance, the pull request titled ``use github.ref rather than github.event.ref in deploy.yml'' of the \texttt{tldraw} GitHub repository\footnote{\url{https://github.com/tldraw/tldraw/pull/2495}} illustrates an instance of CI/CD enhancement. In this case, developers include  Dependabot\footnote{\url{https://github.com/dependabot/dependabot-core}}, a tool to automatically update dependencies, as part of the CI/CD process~\cite{tldraw_2980}. This means that automating the dependency update process comes with a cost: an additional workload, as developers need to set up the new tool in conjunction with the project's workflows. 
This example aligns with previous studies showing that integrating bots to automate dependency updates generates maintenance overhead for developers~\cite{rombaut2023there}.

More generally, previous research has shown that software artifacts similar to workflows, such as build files and bots, can introduce additional maintenance tasks in software projects~\cite{mcintosh2011empirical,rombaut2023there}. This suggests that workflow files, which are also used to automate repetitive tasks in professional software production, may generate extra workload for developers. However, the nature of such effort has not been well studied.

We present a large-scale empirical study towards characterizing the effort needed to maintain GA workflow files. We review the change history of 183 projects hosted on GitHub using GA across 10 programming languages. Inspired by previous work on similar kinds of development artifacts~\cite{jiang2015co,mcintosh2011empirical}, our study focuses on three primary aspects: (1) the number and size of workflow files distributed across multiple projects as well as the frequency of changes (\ie file churn rate), 
(2) workflow coupling, which denotes how frequently source code changes require workflow changes, 
and (3) workflow ownership, which involves identifying the developers responsible for these changes.

On the one hand, as expected, we find that workflow files constitute a rather small proportion of a project's total amount of files, and that they are changed less frequently than other kinds of source code files.
On the other hand, however, the average size of workflow files is comparable to those of production and test code files, and they are far from being stable in terms of evolution. 
Developers tend to update production and test code when workflow files are modified, primarily due to bug fixing and CI/CD improvement. 
Changes are often conducted by only a few developers, but these developers also take the roles of production and test code developers.

The implications of our study are manifold. 
Practitioners should be aware that while automation through GA workflows helps to streamline the CI/CD process~\cite{wessel2023github}, it also incurs significant maintenance efforts expressed in workflow modifications. 
This awareness can lead to better planning and resource allocation for maintaining GA automation workflows. 
Moreover, like any other artifact in professional software production, GA workflows should be designed with maintainability in mind.
From a researchers' perspective, our findings highlight the need to identify and document best practices for maintaining GA workflows. In addition, supporting tools should be equipped with features that facilitate easier updates and maintenance, particularly regarding dependency tracking and improved error reporting.

In summary, this paper provides the following contributions:  

\begin{itemize}
\item A large-scale empirical study providing quantitative results on the volume, evolution and ownership of GA workflow files as well as their logical coupling with other source code files.
\item A detailed qualitative investigation that yields a taxonomy comprising four major reasons for the logical coupling of workflow files and other source code files.
\item A dataset curated from 183 mature GitHub projects facilitating further research on studying the evolution and maintenance of GA workflows.
\end{itemize}

The data and tools used in our study are publicly available for the sake of replication and reproduction~\cite{valerp}.

\section{Research Questions}

The goal of this study is to characterize the maintenance of GA workflow files. We propose five concrete research questions for guiding our study, inspired by prior empirical research on the maintenance of similar artifacts~\cite{mcintosh2011empirical,jiang2015co}. With these RQs, we aim to better understand (i) the {\em volume and evolution} of workflow files within a software project (RQ\textsubscript{1} and RQ\textsubscript{2}), (ii) the {\em logical coupling} of workflow files and other source code files including the contributing factors (RQ\textsubscript{3} and RQ\textsubscript{4}), and (iii) {\em workflow ownership} in terms of developers being responsible for changing workflow files (RQ\textsubscript{5}):

\setlength{\leftmargini}{0pt}

\noindent \textbf{RQ\textsubscript{1}: How many workflow files are in a project, and what are their sizes?} We aim to quantify the volume of workflow files to assess their significance across multiple projects. Determining their size serves as first high-level indicator for assessing their complexity.

\noindent \textbf{RQ\textsubscript{2}: What is the rate of changes in workflow files?} We seek to analyze the rate of change in workflow files (\ie file churn rate~\cite{nagappan2007using}) by measuring the fraction of workflow files that have undergone changes over a specified period. This is motivated by previous research by Shin \etal, and Jiang and Adams~\cite{shin2010evaluating, jiang2015co}, which underscores that a high churn rate may indicate the presence of flawed or potentially vulnerable files, requiring increased maintenance effort. 

\noindent  \textbf{RQ\textsubscript{3}: To what extent are workflow files logically coupled with other source code files?} We want to examine the degree of interdependence between workflow files and other file categories using logical coupling~\cite{gall1998detection}. 
Previous studies indicate that artifacts similar to workflows, such as build and Infrastructure as Code files, exhibit low coupling with source code files~\cite{jiang2015co,mcintosh2011empirical}. In this context, we aim to determine whether this is different for workflow files or holds true for them as well.

\noindent \textbf{RQ\textsubscript{4}: What are the reasons that contribute to workflow files being logically coupled with other source code files?} Given that some degree of logical coupling can be observed, we aim to uncover the causes of logical coupling between workflow files and other file categories.
By knowing the reasons behind such coupling across multiple projects, we can assess whether it is an inherent aspect of the system architecture or if there is a common pattern of developers' behavior. We identify these reasons through a qualitative analysis of logical workflow coupling and develop a comprehensive taxonomy of the contributing factors.

\noindent \textbf{RQ\textsubscript{5}: How is workflow ownership distributed among different roles of developers?} We want to determine how many developers are involved in maintaining workflow files, and how such ownership is distributed among different roles of developers. This will help us to understand how responsibilities are distributed within developer teams and whether there are potential bottlenecks or knowledge silos.

\section{Study Design}


In this section, we give an overview of our study design comprising three major phases, namely project selection, data curation, and data analysis. We structure Sections~\ref{sec:PS} and \ref{sec:DC} according to the individual steps of each of the first two phases. For the sake of avoiding redundant descriptions, Section~\ref{sec:AT} provides a more general introduction into the analysis techniques used in our study, as they are partially reused across our five RQs. We will describe their applications in terms of the concrete analyses in more detail in Section~\ref{sec:results}. 

\subsection{Project Selection} \label{sec:PS}

We collected a dataset from GitHub projects using GA. Our records include source code change logs and details about the developers responsible for the changes.
To ensure the quality and relevance of our dataset, we followed the methodology proposed by Kalliamvakou \etal~\cite{kalliamvakou2014promises}, splitting the selection process into four inclusion criteria: 
(a) initial inclusion criteria, 
(b) projects that use GA and specific programming languages, 
(c) active, and large projects, and 
(d) software projects that are not duplicates of each other.

\emph{(a) Initial inclusion criteria:} We leverage the \texttt{seart-ghs} website~\cite{Dabic:msr2021data} to identify software projects on GitHub~\cite{Dabic:msr2021data}. To find projects with a rich change history, we selected only those with at least 500 commits. 
Furthermore, using the number of stars as a proxy for a project's popularity~\cite{github_stars}, we only included projects with at least 500 stars to avoid irrelevant or toy projects.
We did not include forked projects because they largely contain duplicated project histories, which would bias our analysis. To include a large number of projects, we selected those created before December 31, 2023.

\emph{(b) Projects using popular programming languages and GA:} We considered projects from the top programming languages from 2020 to 2023 (namely JavaScript, Python, TypeScript, Java, C\#, C++, PHP, C, Shell, and Ruby) due to their relevance in the GitHub community during the study period~\cite{githubblog2023}. We identified projects using GA through the GitHub API because they are the focus of our study.

\emph{(c) Active and large projects:} We aimed to include those projects demonstrating a collaborative, long-term software development process. To identify such projects, we defined thresholds using the ``knee method'', following the strategy outlined by Weeraddana \etal~\cite{weeraddana2023empirical}. This strategy involved evaluating: (1) the number of commits, (2) the number of contributors, and (3) the number of stars. To evaluate the number of commits, we established a threshold of 8.151, chosen due to its proximity to the ``knee'' of the curve, identifying 6,867 projects. For the number of contributors, a threshold of 135 was similarly selected, also close to the curve's knee, identifying 2,417 projects. 
Likewise, for the number of stars, we set a threshold of 28,691, identifying 211 projects with significant popularity.
From these projects, we only selected those  providing a ``readme'' file written in English, resulting in 187 projects.

\emph{(d) Deduplication of software projects:} Finally, we manually identified and removed duplicate and non-software projects before cloning the remaining ones. 
We found that the projects \texttt{AutoGPT} and \texttt{Auto-GPT} are the same non-forked repository from the \texttt{Significant-Gravitas} owner, so we excluded the second. After excluding three more projects due to cloning errors, our dataset consists of 183 diverse software projects from well-known organizations such as Google, Microsoft, and others.

\subsection{Data Curation} \label{sec:DC}


The data curation process consists of four steps: 
(a) Classifying projects into single-workflow and multi-workflow groups, 
(b) collecting commit data for each project, 
(c) classifying a project's files into production code, test code, and workflow files, and 
(d) unifying contributor identities.

{\em (a) Classification into single- and multi-workflow projects:}
To account for different strategies of workflow modularization, we classify the projects into two groups: single-workflow, consisting of 27 projects with only one workflow file, and multi-workflow, containing 156 projects with more than one workflow file. Our study compares these groups to assess differences in their maintenance and evolution characteristics.

{\em (b) Collecting commit data for each project:}
We collect commit data for each project by iterating through the commit history of all branches using the command \texttt{git log --topo-order}. We extract the unique commit ID (SHA), timestamp, contributor's name, and the files modified or created for each commit. We filter the commits to include only those made from the point of introduction of the first GA-related commit up to the end of 2023. This ensures that the timeframes will be associated with the use of GA workflows.

{\em (c) Unifying contributor identities:} 
Changes in a committer's name or email address can result in incorrect attribution of commits on GitHub. To unify contributor identities, we use the GitHub-alias-merging script by Vasilescu \etal~\cite{vasilescu2015data}, which employs heuristics to link different aliases and email addresses of the same contributor.

{\em (d) Classifying a project's files into production code, test code, and workflow files:}
We classify the files in each project as production, test, or workflow files, following the method of Nejati \etal~\cite{nejati2023icse}. We disregard any files that do not fall into these categories. We use file extensions, naming, and location conventions to identify file types.  
We focus on source code files written in the project's primary programming language, categorizing them as production or test code files. For example, in Java projects like \texttt{spring-boot}\footnote{\url{https://github.com/spring-projects/spring-boot}} and \texttt{dbeaver}\footnote{\url{https://github.com/dbeaver/dbeaver}}, we consider files with the \texttt{.java} extension. Using project-specific naming and location conventions, we create regular expressions to identify test files, classifying non-matching source files as production code. We identify workflow files by their location in the \texttt{.github/workflows} directory and their \texttt{.yml} or \texttt{.yaml} extension. Unlike previous studies, we exclude build files and similar artifacts due to the complexity of identifying these across multiple programming languages (e.g., due to the heterogeneity of build systems and the various files involved as build scripts).

\begin{table*}[t]
\centering
\caption{Association Rule metrics: Support, Confidence and Lift~\cite{agrawal1993mining} (here: to assess the coupling relationship between changes in different file categories).}
\begin{tabular}{>{\centering\arraybackslash}m{0.18\columnwidth}>{\centering\arraybackslash}m{0.14\columnwidth}>{\raggedright\arraybackslash}m{1.6\columnwidth}}
\toprule
\textbf{Metric} & \textbf{Formula} & \textbf{Description} \\
\midrule
Supp(A) & $P(A)$ & Determines the frequency of changes in file category A. A value of 1 (or 100\%) means that every commit changes at least one file in category A. \\
\midrule
Supp(A \& B) & $P(A \cap B)$ & Determines how frequently changes in file categories A and B occur together. A maximum value of 1 means that every commit changes files in categories A and B. \\
\midrule
Conf(A$\Rightarrow$B) & $\frac{P(A \cap B)}{P(A)}$ & Determines how often changes in file category B occur along with changes in file category A. A maximum value of 1 means that whenever a file of category A is changed, at least one file of category B is changed in the same commit. \\
\midrule
Lift(A$\Rightarrow$B) & $\frac{P(A \cap B)}{P(A)P(B)}$ & Determines how much more (or less) likely changes in file category B are to occur along with changes in file category A, compared to when no file of category A has changed. Lift = 1 means that changes in file categories A and B occur together as expected if they were independent, while Lift \textgreater 1 (Lift \textless 1) indicates a strong (weak) association A $\Rightarrow$ B. \\
\bottomrule
\end{tabular}
\label{tab:ar_metrics}
\vspace{-5mm}
\end{table*}

\subsection{Analysis Techniques} \label{sec:AT}


We use three main analysis techniques: 
(a) statistical tests, 
(b) association rules, and
(c) open coding \& card sorting.

\emph{(a) Statistical tests:} We primarily used the Kruskal-Wallis and Mann-Whitney tests~\cite{scipy-kruskal} for our statistical analyses (RQ\textsubscript{1}-RQ\textsubscript{5}). We applied the non-parametric Kruskal-Wallis test to determine if the distribution of a measure varied between the three file categories. If we rejected the null hypothesis—stating \textit{``there is no significant difference between the means of the three categories''}—it indicated that at least one category had a distinct distribution of the metric under examination. For post-hoc testing, we used Mann-Whitney tests to identify which specific categories had distinct distributions. We performed these tests between every pair of file categories, applying the Bonferroni correction to the alpha value (0.05 by default in all our tests).

\begin{figure}[t]
    \centering
    \includegraphics[scale=0.57]{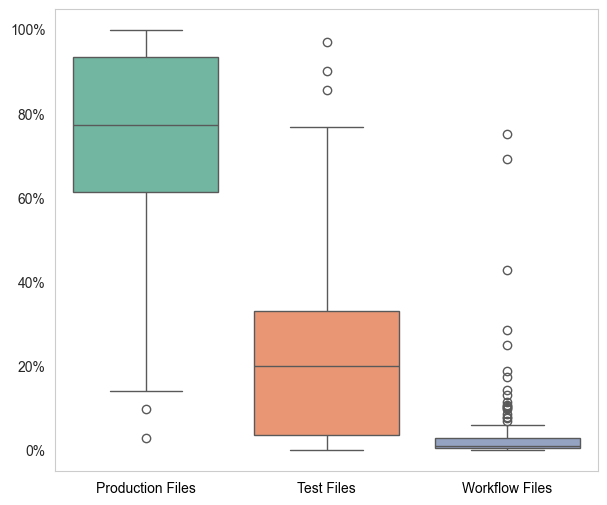}
    \caption{Mean relative frequencies of the studied file categories in the multi-workflow group.}
    \label{fg:rq1_a_multi}
    \vspace{-5mm}
\end{figure}

\emph{(b) Association rules:} We investigate logical coupling in RQ\textsubscript{3} and RQ\textsubscript{5} through association rules~\cite{agrawal1993mining}. 
Specifically, to assess the coupling relationship between different file categories, we analyze each pair \(\langle A, B \rangle\) of file categories. 
As for quantitative assessment, we use the usual metrics known as \emph{Support (Supp)}, \emph{Confidence (Conf)} and \emph{Lift}, as defined in Table \ref{tab:ar_metrics} for assessing frequencies of and relationships between file category changes per commit (RQ\textsubscript{3}). The metrics used for assessing frequencies of and relationships between developer roles per project history (RQ\textsubscript{5}) are defined analogously.
Afterwards, similar to Jiang and Adams~\cite{jiang2015co}, we performed chi-square statistical tests to test whether the obtained confidence values were significant or not higher than expected due to chance.

\emph{(c) Open coding \& card sorting:} For RQ\textsubscript{4}, we conduct open coding~\cite{charmaz2014grounded} using a randomly selected set of logically coupled commits that have a lift greater than one. Then, we apply open card sorting~\cite{zimmermann2016card} to codify the reasons for coupled changes. Open Coding and Card Sorting are widely used techniques in software engineering useful to derive taxonomies from data~\cite{zimmermann2016card}.

\section{Analysis and Results}
\label{sec:results}

\subsection{Relative Frequency and Size of Workflow Files (RQ\textsubscript{1})} 

To answer RQ\textsubscript{1}, we study the relative frequency and size of each file category (i.e., production, test, and workflow) across all projects, separated by single-workflow and multi-workflow groups. We use the Kruskal-Wallis test to evaluate the statistical significance of our findings.  

\begin{figure}[t]
    \centering
    \includegraphics[scale=0.58]{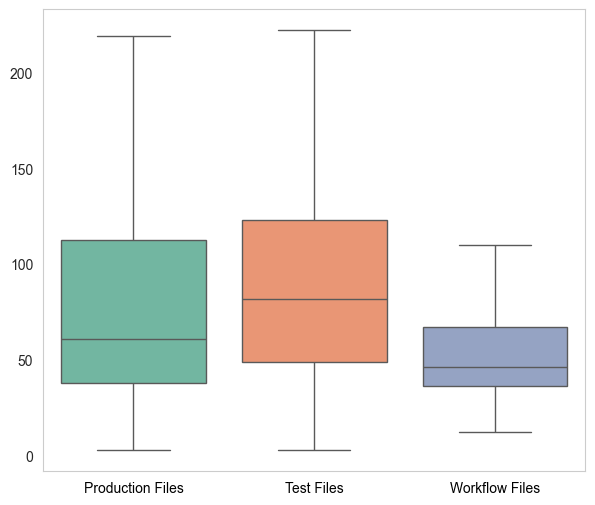}
    \caption{Mean file sizes in NLOC (without outliers) of the studied file categories in the multi-workflow group.}
    \label{fg:rq1_b_multi}
    \vspace{-5mm}   
\end{figure}

Overall, the 183 studied projects contain between 1 and 70 workflow files. The multi-workflow group has a mean relative frequency of 3.74\% workflow files, while the single-workflow group has 3.05\%. Production files make up the most significant proportion, with a mean relative frequency of 74.59\% in the multi-workflow group and 83.31\% in the single-workflow group. Test files are the second-largest category, with a mean relative frequency of 21.65\% in the multi-workflow group and 13.63\% in the multi-workflow group. Figure~\ref{fg:rq1_a_multi} shows a boxplot of the proportion of the three file categories in the multi-workflow group, reflecting these trends. We do not include a boxplot of the single-workflow group since it follows the same trends. More detailed statistics on both distributions may be found in our replication package.

As we are dealing with textual source code artifacts in our study, we measure their size in Number of Lines of Code (NLOC). Production and test files are generally larger than workflow files. In the multi-workflow group, workflow files have a mean NLOC of 60.99, while production and test files expose mean NLOC values of 103.26 and 101.37, respectively. In the single-workflow group, the mean NLOC increases to 98.93 for workflow files, 128.19 for production files, and 113.32 for test files. Figure~\ref{fg:rq1_b_multi} shows a boxplot of the mean file sizes in NLOC (without outliers) of the studied file categories in the multi-workflow group. We do not include a boxplot for the single-workflow group since it follows the same trends. Again, more detailed statistics on both distributions may be found in our replication package.

\begin{tcolorbox}[colback=blue!5!white, colframe=blue!5!white]
\textbf{Finding 1:} Workflow files account for a small proportion of files, with a mean relative frequency of 3.74\% in the multi-workflow group and 3.05\% in the single-workflow group. However, their mean size of NLOC, at 60.99 in the multi-workflow group and 98.92 in the single-workflow group, is close to that of production and test files, which are of the same order of magnitude.
\end{tcolorbox}

\subsection{Rate of Changes in Workflow Files (RQ\textsubscript{2})} 
To assess the rate of changes in workflow files, we first determined the amount of changed files per month of each project. Subsequently, to enable comparisons over time, we normalize the number of changed files by dividing it by the total number of files in the corresponding category for that month, yielding the proportion of changed files in terms of file churn rate~\cite{nagappan2007using}. For each project, we then calculate the average proportion of changed files per month, and we study the distribution of this average across all projects. Furthermore, we conducted a Kruskal-Wallis test and post-hoc tests to examine the statistical significance of our results.

The analysis of both single- and multi-workflow groups reveals that the monthly proportion of changed workflow files is significantly lower than for production and test code files. In the single-workflow group, workflow files have a mean value of 6.8\%, while in the multi-workflow group, the mean value is 8.2\%. Production files have mean values of 84.53\% and 74.12\% for the single-workflow and multi-workflow groups, respectively. Test code files have mean values of 10.75\% and 18.90\%, correspondingly. 

\begin{figure}[t]
    \centering
    \includegraphics[scale=0.6]{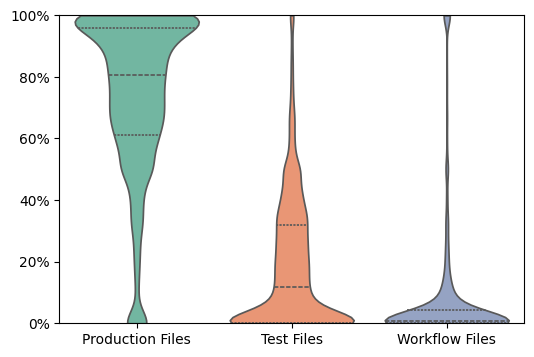}
    \caption{Distribution of the rate of changes for the workflow, test, and production file categories. We calculated the monthly rate of change by normalizing the number of changed files against the total files in each category, then averaging this proportion per month for each project.}
    \label{fg:rq2_multi_a}
    \vspace{-5mm}
\end{figure}

Figure~\ref{fg:rq2_multi_a} illustrates the distribution of the rate of changes for the workflow, test, and production file categories. The solid lines represent the median percentages, and the dotted lines represent the first and third quartiles. The figure shows that workflow files have a relatively narrow distribution with a concentration around lower change rates, indicating they experience fewer changes overall. Additionally, the plot for workflow files reveals a single peak, suggesting a consistent rate of changes, mainly concentrated at lower proportions. In contrast, production files show a wide distribution with a higher concentration around the upper change rates and a pronounced peak, reflecting their more frequent updates. Test files exhibit a distribution between the workflow and production files, indicating moderate change frequencies.

A Kruskal-Wallis test confirmed significant differences in the distributions of the monthly change percentages among the different file categories in both groups, with extremely low p-values of 5.06e-33 for the multi-workflow group and 0.0 for the multi-workflow group. Further Mann-Whitney post-hoc tests showed statistically significant differences between Production and Workflow files, with p-values of 1.77e-14 and 0.0, respectively. The tests also showed significant differences between Production and Test files, with p-values of 2.04e-29 and 0.0 each. Additionally, there were significant differences between Test and Workflow files, with p-values of 4.76e-11 for the multi-workflow group and 1.32e-255 for the multi-workflow group. 

\begin{tcolorbox}[colback=blue!5!white, colframe=blue!5!white]
\textbf{Finding 2:} Workflow files change less frequently than production and test code files. In the multi-workflow group, a mean of 8.2\% of the workflow files undergo at least one monthly change, while repositories with a single-workflow account for a mean 6.9\%. 
In the same time period, production files have a mean change proportion of 74.12\% in the multi-workflow group and 84.53\% in the single-workflow group. Similarly, test files change 18.90\% of the time in multi-workflow and 10.75\% of the time in repositories with a single workflow.
\end{tcolorbox}

\subsection{Quantitative Analysis of Logical Workflow Coupling (RQ\textsubscript{3})} 

The results of our quantitative analysis indicate that workflow files are the least frequently changed, as evidenced by the low \emph{Support} metric values (1.40\% in the multi-workflow group and 3.13\% in the single-workflow group)(See Table~\ref{tab:rq3}). While modified more often than workflow files, test files still show relatively low-frequency changes (\emph{Support} values of 19.07\% in the multi-workflow group and 32.57\% in the single-workflow group). On the other hand, production files exhibit the highest frequency of changes, with \emph{Support} values of 92.57\% in the multi-workflow group and 88.55\% in the single-workflow group.

Regarding the relationship between workflow and test files (\emph{Supp(Workflow \& Test)}), the \emph{Support} is 0.06\% in the single-workflow group and 0.23\% in the multi-workflow group, indicating that commits involving changes in both categories are extremely rare. For the relationship between workflow and production files (\emph{Supp(Workflow \& Production)}), the \emph{Support} is slightly higher, with values of 0.17\% in the single-workflow group and 0.53\% in the multi-workflow group. 

Changes in workflow files are rarely associated with changes in other categories, and when they are, they are more likely to coincide with changes in production files rather than test files. Regarding the relationship between workflow and test files (\emph{Conf(Workflow $\Rightarrow$ Test)}), the \emph{Confidence} is 4.60\% in the multi-workflow group and 7.40\% in the single-workflow group, indicating that if a commit involves changes in workflow files, there is a 4.60\% chance in the multi-workflow group and a 7.40\% chance in the single-workflow group that it also involves changes in test files. For the relationship between workflow and production files (\emph{Conf(Workflow $\Rightarrow$ Production)}), the \emph{Confidence} values are higher, with 12.26\% in the single-workflow group and 17.00\% in the multi-workflow group. If a commit involves changes in workflow files, there is a 12.26\% chance in the single-workflow group and a 17.00\% chance in the multi-workflow group that it also involves changes in production files.


Changes in workflow files tend to occur independently of changes in both test and production files, as indicated by the \emph{Lift} metrics. For the relationship between workflow and test files (\emph{Lift(Workflow $\Rightarrow$ Test)}), the \emph{Lift} is 24.12\% in the single-workflow group and 22.71\% in the multi-workflow group. These values indicate that changes in workflow files reduce the likelihood of changes in test files to 24.12\% and 22.71\% of what would be expected by random chance, respectively. This suggests a negative association, meaning that changes in workflow files rarely coincide with changes in test files.

For the relationship between workflow and production files (\emph{Lift(Workflow $\Rightarrow$ Production)}), the \emph{Lift} values are 13.25\% in the single-workflow group and 19.20\% in the multi-workflow group. These values show that when workflow files change, production files rarely change at the same time: only 13.25\% of the time in the single-workflow group and 19.20\% of the time in the multi-workflow group. Although the \emph{Lift} values for workflow and production files are slightly higher than those for workflow and test files, they still indicate a negative association.


The combination of workflow and test yields a chi-square value of 4502.59 with a p-value of $<$ 0.0001, suggesting a strong correlation between these two categories. Furthermore, the combination of workflow and production reveals an even stronger association, with a chi-square value of 78,799.33 and a p-value of $<$ 0.0001. Similarly, the combination of test and production also shows a highly significant correlation, with a chi-square value of 58,511.84 and a p-value of $<$ 0.0001. These results imply that there are significant relationships between these categories and that they are not independent of each other.

\begin{table}[t]
\centering
\caption{File category changes per commit: Support, Confidence, and Lift metrics for single- and multi-workflow groups.}

\begin{tabular}{llrr}
\toprule
Metric & Category Combination & Single & Multi \\
\midrule
\multirow{5}{*}{Support} & Workflow & 0.0140 & 0.0313 \\
 & Test & 0.1907 & 0.3257 \\
 & Production & 0.9257 & 0.8855 \\
 & Workflow \& Test & 0.0006 & 0.0023  \\
 & Workflow \& Production & 0.0017 & 0.0053  \\
\midrule
\multirow{2}{*}{Confidence} 
 & Workflow $\Rightarrow$ Test & 0.0460 & 0.0740 \\
 & Workflow $\Rightarrow$ Production & 0.1226 & 0.1700 \\
\midrule
\multirow{2}{*}{Lift} & Workflow $\Rightarrow$ Test & 0.2412 & 0.2271 \\
 & Workflow $\Rightarrow$ Production & 0.1325 & 0.1920\\
\bottomrule
\end{tabular}
\label{tab:rq3}
\vspace{-5mm}
\end{table}

\begin{tcolorbox}[colback=blue!5!white, colframe=blue!5!white]
\textbf{Finding 3:} The coupling between workflow files and other file categories is minimal and weak. For the relationship between workflow and test files, there is a 4.60\% chance in the single-workflow group and a 7.40\% chance in the multi-workflow group that a commit involving workflow files also changes test files. For workflow and production files, there is a 12.26\% chance in the single-workflow group and a 17.00\% chance in the multi-workflow group. In the multi-workflow group, there is a slightly higher but still weak interdependence between workflow and production files.
\end{tcolorbox}

\subsection{Qualitative Analysis of Logical Workflow Coupling (RQ\textsubscript{4})}

\begin{table*}[ht]
\centering
\caption{Main Reasons for Logical Coupling.}
\begin{tabular}{llcc}
\toprule
\textbf{Main Reasons for Logical Coupling} & \textbf{Specific reason} & \textbf{Workflow \& Test} & \textbf{Workflow \& Production} \\
\midrule

\multirow{3}{*}{Management and Configuration} 
 & Configuration and Infrastructure &  1\% & 3\% \\
 & Dependency and Build Management & 7\% & 6\% \\
 & Merging and Synchronization & 1\% & 5\% \\
\midrule
\multirow{3}{*}{Development and Enhancement} 
 & CI/CD Improvement & 61\% & 33\% \\
 & Code Quality and Maintenance & 7\% & 9\% \\
 & Features and Enhancements &  6\% & 11\% \\
\midrule
\multirow{1}{*}{Bug Fixes and Compatibility} 
 & --- & 12\% & 18\% \\
\midrule
\multirow{2}{*}{Testing and Documentation} 
 & Testing & 4\% & 12\% \\
 & Documentation & 1\% & 3\% \\
\midrule
\multirow{1}{*}{Total} 
 &   & 100\% & 100\% \\
\bottomrule
\end{tabular}
\label{tab:rq4}
\vspace{-2mm}
\end{table*}

In addition to RQ\textsubscript{3}, we also conducted a qualitative analysis of commits that contribute to the highest levels of logical coupling, as indicated by their Lift values (see Table~\ref{tab:rq3}). We applied open coding and card sorting for this analysis. We selected 200 commits with the highest Lift scores (100 from the single-workflow group and 100 from the multi-workflow group) to investigate why workflow changes were tightly coupled, similar to the previous work by Jiang et al.~\cite{jiang2015co}.

We followed a structured approach to card sorting, carried out in three distinct phases as recommended by Zimmermann~\cite{zimmermann2016card}: 
First, in the {\em preparation phase} we use descriptive coding~\cite{lo2023cognitive}. We set up a collection of cards to provide a detailed understanding of the workflow coupling. Each commit was meticulously described based on the GitHub single commit view, including the reasons for and locations of the changes. 
Second, in the {\em execution phase}, we labelled and organized each card into relevant groups with clear, descriptive categories. 
Finally, in the {\em analysis phase}, we created abstract hierarchies without predefined categories to identify broader trends and general categories.

To ensure accuracy, the description criteria were tested by three participants: two undergraduate students, one graduate student and the first author. Each of them was given 100 commits to describe the reason for coupling. These descriptions were then reviewed and refined collaboratively to establish a final structure. During the sorting process, we used spreadsheets to facilitate easier handling and better understanding of each commit. A hierarchy was created manually to organize a set of categories.

As a result, we built a taxonomy consisting of nine reasons for logical coupling including a total of 200 labelled commits (Table~\ref{tab:rq4}). 
We grouped them into four main categories: 
(1) Management and Configuration, 
(2) Development and Enhancement, 
(3) Bug Fixes and Compatibility, and 
(4) Testing and Documentation.

\subsubsection{Management and Configuration}  This category involves tasks related to setting up and managing the project's environment, including configuring infrastructure, handling dependencies, and synchronizing code across different branches (9\% of workflow \& test, 13\% of workflow \& production). These activities ensure that the foundational aspects of the project are well-organized and functioning correctly. Three reasons for coupling are related to this category: (i) Configuration and Infrastructure, (ii) Dependency and Build Management, and (iii) Merging and Synchronization.

\subsubsubsection{Configuration and Infrastructure} encompasses modifications related to setting up and maintaining the infrastructure and configuration necessary for the project. Proper configuration management is essential for ensuring a stable and efficient development environment. The commit ``fix(dev): Add .python-version back'' illustrates this by addressing issues related to the Python version management within the development environment.\footnote{\url{https://bit.ly/3RDUhR8}}

\subsubsubsection{Dependency and Build Management} involves managing the project's dependencies, orchestrating the build process, and ensuring that all necessary components are correctly integrated. The commit ``Revert Switch to Debian 11 (bullseye) as base for our dockerfiles'' involves reverting a change to Debian 11 (bullseye) that caused all pull requests to fail.\footnote{\url{https://bit.ly/3VEq99g}}
The revert restored stability, ensuring dependencies and configurations returned to a reliable state.

\subsubsubsection{Merging and Synchronization} cover tasks related to merging code from different branches and synchronizing changes across the codebase. The commit ``Merge branch `6.4' into 7.0'' exemplifies this category by integrating changes from branch 6.4 into 7.0, which involves combining contributions from different parents and resolving potential conflicts.\footnote{\url{https://bit.ly/45AHXqG}}
It addresses specific issues, such as fixing the silencing of the wait command for \texttt{sigchild-enabled} binaries, and updates multiple files to ensure consistent behaviour and compatibility. 

\subsubsection{Development and Enhancement}

This category encompasses tasks related to improving the project's functionality and performance. It involves introducing new features, enhancing existing ones, and ensuring the software efficiently meets user needs and standards (84\% of workflow \& test, 53\% of workflow \& production). Three factors contributing to coupling are associated with this category: (i) CI/CD Improvement, (ii) Code Quality and Maintenance, and (iii) Features and Enhancements.

\subsubsubsection{CI/CD Improvement}  relates to optimizing and automating the continuous CI/CD Improvement. It includes setting up and refining pipelines and ensuring smooth and efficient code integration. The commit ``Use npm v7 with workspaces for local development and testing'' explains how developers upgrade the project to use \texttt{npm} v7 with  \texttt{workspaces}, enhancing continuous integration and deployment processess.\footnote{\url{https://bit.ly/3RBzSMw}}

\subsubsubsection{Code Quality and Maintenance} involve activities that ensure the software remains functional, efficient, and easily understood over time. This includes practices such as refactoring code to improve readability and performance and writing and maintaining tests to catch bugs early. The commit ``tools: automate histogram update'' enhances the CI/CD workflow by automating the update process for the project's histogram dependency.\footnote{\url{https://bit.ly/3VBIaFl}}
It adds a new script, update-\texttt{histogram.sh}, to the \texttt{tools} directory, which fetches and updates the histogram to its latest version from GitHub releases. 

\subsubsubsection{Features and Enhancements} involve activities to add new functionality or improve existing features in a software project. This includes designing and implementing new features to meet user needs, optimizing performance, enhancing user interfaces, and extending the software's capabilities. For example, the commit ``Add --os and --arch flags to readall'' exemplifies this category by adding the \texttt{--os} and \texttt{--arch} flags to the \texttt{brew readall} command, allowing users to read using specified operating systems and CPU architectures.\footnote{\url{https://bit.ly/4eyYX4E}}

\subsubsection{Bug Fixes and Compatibility}

This category refers to activities dedicated to identifying and fixing bugs, as well as ensuring that the project remains compatible with different environments and dependencies. These tasks are crucial for maintaining the stability and reliability of the project (12\% of Workflow \& Test, 18\% of Workflow \& Production). The commit ``alt-svc: enable by default'', illustrates this category: \texttt{altsvc} support by default in curl and removes the unused \texttt{CURLALTSV CIMMEDIATELY} option.\footnote{\url{https://bit.ly/4cuZ4fK}}
The update involves modifications across 27 files, including changes to configuration files for CI pipelines, such as \texttt{.azure-pipelines.yml}, and \texttt{.github/workflows/macos.yml}. 

\subsubsection{Testing and Documentation} This category focuses on tasks related to ensuring the reliability of the code through testing and improving project documentation (5\% of Workflow \& Test, 15\% of Workflow \& Production). Two factors contributing to coupling are associated with this category: (i) Testing, and (ii) Documentation.

\subsubsubsection{Testing} refers to modifications that enhance the testing framework or add new tests to ensure more comprehensive and effective software validation. This includes updates to testing scripts, configurations, or the addition of new test cases. The commit titled ``fix: nschematics install Windows'' addresses issues with installing the schematics on Windows systems.\footnote{\url{https://bit.ly/4exe8LJ}}
The changes include adding smoke tests for Angular Schematics on multiple operating systems (Ubuntu, Windows, macOS), updating \texttt{Node.js} setup in GA, and modifying various files to support the Windows installation process.

\begin{table}[t]
\centering
\caption{Developer roles per project history: Support and Confidence metrics for single- and multi-workflow groups.}
\begin{tabular}{llrr}
\toprule
Metric & Category Combination & Single & Multi \\
\midrule
\multirow{6}{*}{Support} 
 & Workflow & 0.0435 & 0.0486 \\
 & Test & 0.2150 & 0.4664 \\
 & Production & 0.9746 & 0.9636 \\
 & Workflow \& Test & 0.0165 & 0.0224 \\
 & Workflow \& Production & 0.0155 & 0.0225 \\
 & Test \& Production & 0.1751 & 0.4156 \\
\midrule
\multirow{4}{*}{Confidence} 
 & Workflow $\Rightarrow$ Test & 0.3809 & 0.4609 \\
 & Workflow $\Rightarrow$ Production & 0.3571 & 0.4637 \\
 & Test $\Rightarrow$ Workflow & 0.0771 & 0.0480 \\ 
 & Production $\Rightarrow$ Workflow & 0.0159 & 0.0233 \\ 
\bottomrule
\end{tabular}
\label{tab:rq5}
\vspace{-3mm}
\end{table}

\subsubsubsection{Documentation} refers to the changes to the documentation within the software project. This includes updates, additions, or deletions to ensure that the documentation is accurate and up-to-date. The commit ``Change supported PyPy versions to 3.9 and 3.10'', illustrates the updates the the ReadTheDocs (RTD) URL for the coverage documentation is updated to point to the latest version.\footnote{\url{https://bit.ly/3xsxYHi}}
There are also adjustments to the GA workflow, specifically for testing on \texttt{PyPy} versions and ensuring the correct system libraries are installed based on the Python version being used. 

\begin{tcolorbox}[colback=blue!5!white, colframe=blue!5!white]
\textbf{Finding 4:}  Given that some degree of logical coupling of workflow files and other file categories can be observed, our results identified key reasons for this. The main contributors are \emph{Management and Configuration} tasks, with significant activities including \emph{Dependency and Build Management}, and \emph{Merging and Synchronization}. The primary drivers are \emph{Development and Enhancement} activities, particularly \emph{CI/CD Improvement} processes. \emph{Bug Fixes and  Compatibility} also play a critical role, along with \emph{Testing and Documentation} efforts.
\end{tcolorbox}

\subsection{Workflow Ownership Distribution (RQ\textsubscript{5})} 

Identifying ownership for each category involves examining the commits made to the repository. We look at the author of each commit to determine their role in the development process. If a commit modifies a workflow file, then the author of that commit is considered a workflow developer. An author can have multiple roles, such as being both a workflow developer and a production code developer, even for the same commit.

We compute the same metrics as for RQ\textsubscript{3}, this time for the  ownership. For example, \emph{Supp(Workflow)} indicates the percentage of developers changing workflow files from the total number of developers. \emph{Supp(Workflow \& Production)} is the percentage of developers changing workflow and production files from the total number of developers. The \emph{Conf(Workflow $\Rightarrow$ Production)} represents the percentage of workflow developers who also modify production files, in relation to the total number of developers who have made at least one change to a workflow file.

Based on the \emph{Support} results, workflow developers have the lowest proportion among all developers, with only 4.35\% in the single-workflow group and 4.86\% in the multi-workflow group being involved in changes to workflow files (Table \ref{tab:rq5}). These results could indicate that workflow changes are more specialized tasks handled by a smaller subset of developers. In contrast, production code developers are the most common among all developers, with 97.46\% in the single-workflow group and 96.36\% in the multi-workflow group making changes to production files. 

Regarding the combination of workflow and test file changes, 1.65\% of developers in the single-workflow group and 2.24\% in the multi-workflow group are involved in both. These low results indicate that it is relatively uncommon for developers to work on both workflow and test files. The infrequency of these combined changes suggests that workflow and test tasks are often handled separately, with few developers integrating their efforts across these areas.

Similarly, 1.55\% of developers in the multi-workflow group and 2.25\% in the single-workflow group are involved in the combination of workflow and production file changes. These low percentages hint that it is uncommon for developers to work on both workflow and production files. The rarity of these combined changes suggests that workflow and production tasks are typically managed independently, with limited overlap in developer responsibilities. 

In terms of \emph{Confidence}, 38.09\% of developers in the single-workflow group who changed workflow files also changed test files, compared to 46.09\% in the multi-workflow group \emph{(Conf(Workflow $\Rightarrow$ Test))}. Although it is relatively uncommon for workflow developers to also be involved in test file changes, these confidence values are moderately high in this context. 
The higher confidence in the multi-workflow group indicates that developers in this group are more likely to integrate their workflow changes with test changes, highlighting a more collaborative and thorough development process.

Most developers do not modify workflow files, suggesting a specialization in these tasks. The considerable values of \emph{Conf(Production $\Rightarrow$ Workflow)} and \emph{Conf(Test $\Rightarrow$ Workflow)} indicate that while workflow developers are relatively few, those involved in production and test activities frequently engage with workflow changes. Specifically, 15.90\% of production code developers in the single-workflow group and 23.30\% in the multi-workflow group also change workflow files, while 7.71\% of test developers in the single-workflow group and 4.80\% in the multi-workflow group also change workflow files.

\begin{tcolorbox}[colback=blue!5!white, colframe=blue!5!white]
\textbf{Finding 5:} Few developers work on workflow files, with only 4.35\% in the single-workflow group and 4.86\% in the multi-workflow group. However, a notable proportion of developers who work on workflow files also work on test files (38.09\% in the multi-workflow group and 46.09\% in the multi-workflow group) and production files (35.71\% in the single-workflow group and 46.37\% in the multi-workflow group).
\end{tcolorbox}

\section{Discussion}

In this section, we discuss the findings presented in the previous section and provide a set of practical implications.

\emph{(a) RQ\textsubscript{1} \& RQ\textsubscript{2}.} 
Compared to production and test code files, workflow files represent only a minor fraction of the overall project files, underscoring their specialized role in managing CI/CD pipelines and automation tasks.
Additionally, workflow files change significantly less frequently than production and test files. This characteristic is likely due to their specific and well-defined roles in automating CI/CD processes, which do not change as dynamically as production and test code. Nonetheless, they are far from being stable, which confirms our hypothesis that manual effort is needed for maintaining workflow files over time. 
%
Although workflow files are typically smaller than production code files, their size is not drastically different from test files, especially in the multi-workflow group where the means are relatively close. That is, workflow files being designed to automate specific tasks still contain a substantial number of lines of code. 

\emph{Implication 1:} 
Given workflow files' substantial size and critical role in CI/CD processes, practitioners should allocate resources accordingly. Despite their smaller proportion, these files require careful maintenance. Practitioners should dedicate sufficient time and personnel to ensure the stability and efficiency of workflows, as this directly impacts the overall development and deployment process.

\emph{(b) RQ\textsubscript{3} \& RQ\textsubscript{4}.} Although the confidence of a workflow file change coinciding with a production file change is relatively low, and even lower with test files, workflow files exhibit minimal coupling with these file categories. A thorough investigation of commits that contribute to the logical coupling reveals that they are primarily updated for CI/CD improvements and bug fixes.

\emph{Implication 2:} 
By categorizing the logical coupling between workflow files and other file categories, our taxonomy reveals that  current tools lack robust error reporting and support features. This classification identifies areas where tools fail to meet users' needs, revealing the need for researchers to develop advanced tools to address these gaps.

\emph{(c) RQ\textsubscript{5}.} Our analysis finds that few developers work on workflow files. However, confidence values reveal that a notable proportion of developers who work on workflow files also work on test and production files, suggesting that workflow maintenance requires cross-functional knowledge. 

\emph{Implication 3:} 
Practitioners should be aware of these dependencies and plan their development and maintenance activities to account for the interconnected nature of changes in workflow files. Organizations should ensure that their teams include developers skilled in both workflow management and general software development.

\section{Threats to Validity and Limitations}

\subsection{Internal Validity} 

{\em Selection Bias:} The selection of projects based on criteria like the number of commits and stars may introduce selection bias, potentially not representing the entire population of GA workflow projects. We mitigated this by using a systematic selection methodology and including a diverse set of projects across multiple programming languages.

{\em Measurement Error and Categorization Accuracy:} Errors or inconsistencies in the repositories' metadata, commit messages, or file histories could affect our results. Additionally, misclassification of file changes into production, test, and workflow categories could occur due to naming conventions and file locations. We used established tools and methods for data curation and analysis, and a rigorous methodology including regular expressions and manual checks, to ensure consistency and accuracy.

{\em Confounding Variables and Temporal Effects:} Factors such as developer experience, project size, and complexity could influence our findings. Additionally, development practices and technologies may have evolved over the four-year study period. To address these issues, we generalized our results across multiple projects and languages, reducing the impact of any single variable. We also included a broad time frame to capture trends and changes over time.

{\em Developer Attribution and File Classification:} Unifying contributor identities can be challenging due to aliasing and changes in email addresses or usernames, and errors in classifying files into production, test, and workflow categories can lead to incorrect conclusions. We used heuristics to merge aliases, manual verification for accurate developer attribution, followed a standardized classification scheme, and performed manual checks to verify accuracy.

\subsection{External Validity} 

{\em Platform and Ecosystem Specificity:} Our study focuses exclusively on GA workflows, which may not apply to other CI/CD platforms such as GitLab CI, Travis CI, or Jenkins. Additionally, projects hosted on GitHub may exhibit unique characteristics compared to those on other platforms. To mitigate this, we suggest future research replicate the study across different CI/CD platforms and hosting environments to verify the generalizability of our results.

{\em Project and Domain Diversity:} Although we included a diverse set of projects across various programming languages, certain project types or domains may still be underrepresented. We addressed this by systematically selecting projects from a variety of domains and ensuring broad representation.


\subsection{Construct Validity}

We assume that studying changes is a good proxy to assess the maintenance of a software artifact. This assumption may not always hold true, as even simple and small changes in a workflow file could result from significant cognitive effort. This simplification was necessary because we aimed to measure across multiple repositories, which is feasible when considering changes rather than the broader concept of maintenance. To mitigate, we performed a qualitative analysis of a sample of commits to better understand the context and complexity of the changes. By examining the reasons for changes we aim to provide a more nuanced understanding of the maintenance activities. 

\section{Related Work}
\label{sec:relWork}

Prior empirical studies have focused on various aspects of maintenance and evolution related to Continuous Integration and Continuous Delivery (CI/CD) practices, essential for modern development using tools like Jenkins, Travis, CircleCI, and GA. Jiang and Adams as well as Adams and McIntosh explored CI configurations and build maintenance efforts, highlighting challenges in maintaining scripts and configurations~\cite{jiang2015co, mcintosh2011empirical}. Gallaba \etal provided a comprehensive overview of CI/CD practices based on eight years of data~\cite{gallaba2022lessons}. Studies on the \texttt{Greenkeeper} dependency bot also emphasize the need to evaluate automation tools' benefits and costs~\cite{rombaut2023there}. Zhao et al.~\cite{zhao2017impact} and Cassee et al.~\cite{cassee2020silent} examined how the introduction of the Travis CI tool affected development practices. 
However, none of the above mentioned studies specifically focused on GA. 

Research shows that GA, introduced in 2019, quickly replaced Travis due to its integration and ease of use~\cite{rostami2023usage}. Rostami Mazrae \etal found modifications to be the most common GA workflow changes, revealing hidden maintenance costs and broader implications~\cite{mazrae2023preliminary}.
Delicheh et al.~\cite{delicheh2023preliminary} provide preliminary insights into the dependencies of GA actions.
Decan et al.~\cite{decan2023outdatedness} have studied the extent to which automation workflows are outdated with respect to updated GA actions, concluding that better policies and tooling are needed to keep workflows up-to-date.
Valenzuela-Toledo and Bergel also identified the need for better tooling for GA workflows~\cite{valenzuela2022evolution}. 
The research by Kinsman \etal examined how developers use GA and how activity indicators change after its adoption~\cite{kinsman2021software}. 
Saroar and Nayebi surveyed developers to understand their perceptions of GA, revealing motivations, decision criteria, and challenges~\cite{saroar2023developers}.
Zhang \etal conducted a large-scale empirical study to create a taxonomy of GA-related problems, while Bouzenia and Pradel looked into resource usage and optimization opportunities in GA workflows~\cite{bouzenia2024resource}. 
Other studies, such as Koishybayev \etal~\cite{koishybayev2022characterizing}, explored security risks associated with excessive privileges in GA workflows. The EGAD project~\cite{valenzuela2023egad,DBLP:conf/iwst/Valenzuela-Toledo23} provides a tool environment to explore and analyze GA workflows, contributing to a deeper understanding of their usage and maintenance practices. Cardoen et al.~\cite{cardoen2024dataset} developed an open-source tool designed to extract the commit histories of changes made to workflow files in GitHub repositories, along with a raw dataset that has been collected using their tool.
Despite the relevance of all ot these studies, none of them specifically addressed the characterization of the maintenance of GA workflow files, which is the primary focus of our research.

\section{Conclusion}

Automating tasks with GA has become standard practice in the software industry, enabling developers to automate building, testing, and deployment processes. However, our research reveals that implementing and maintaining these workflow files is not without additional costs. A large-scale empirical study spanning 183 GitHub projects in ten different programming languages identified several key features and challenges in maintaining GA workflow files.

Our findings confirm results from previous studies on the maintenance of similar artifacts such as Build files and Infrastructure as Code. While beneficial for efficiency, automation entails hidden costs that must be adequately managed. Practitioners must plan and allocate sufficient resources for maintaining these workflows, including identifying and documenting best practices.

Despite the strengths of our study, there are threats to construct validity that need to be considered. The scope of our analysis may not cover all aspects of workflow maintenance, and certain contextual factors unique to individual projects may influence the results. To mitigate these threats in future work, we plan to conduct more focused studies on specific aspects of workflow maintenance, exploring different perspectives and contexts. Additionally, we aim to collaborate with practitioners to gather more detailed insights and develop tailored strategies for efficient workflow management.


\bibliography{references.bib}

\begin{thebibliography}{10}
\providecommand{\url}[1]{#1}
\csname url@samestyle\endcsname
\providecommand{\newblock}{\relax}
\providecommand{\bibinfo}[2]{#2}
\providecommand{\BIBentrySTDinterwordspacing}{\spaceskip=0pt\relax}
\providecommand{\BIBentryALTinterwordstretchfactor}{4}
\providecommand{\BIBentryALTinterwordspacing}{\spaceskip=\fontdimen2\font plus
\BIBentryALTinterwordstretchfactor\fontdimen3\font minus \fontdimen4\font\relax}
\providecommand{\BIBforeignlanguage}[2]{{%
\expandafter\ifx\csname l@#1\endcsname\relax
\typeout{** WARNING: IEEEtran.bst: No hyphenation pattern has been}%
\typeout{** loaded for the language `#1'. Using the pattern for}%
\typeout{** the default language instead.}%
\else
\language=\csname l@#1\endcsname
\fi
#2}}
\providecommand{\BIBdecl}{\relax}
\BIBdecl

\bibitem{golzadeh2022rise}
M.~Golzadeh, A.~Decan, and T.~Mens, ``On the rise and fall of {CI} services in {GitHub},'' in \emph{2022 IEEE International Conference on Software Analysis, Evolution and Reengineering (SANER)}.\hskip 1em plus 0.5em minus 0.4em\relax IEEE, 2022, pp. 662--672.

\bibitem{fowler2006continuous}
M.~Fowler and M.~Foemmel, ``Continuous integration,'' 2006.

\bibitem{humble2010continuous}
J.~Humble and D.~Farley, \emph{Continuous delivery: reliable software releases through build, test, and deployment automation}.\hskip 1em plus 0.5em minus 0.4em\relax Pearson Education, 2010.

\bibitem{fitzgerald2017continuous}
B.~Fitzgerald and K.-J. Stol, ``Continuous software engineering: A roadmap and agenda,'' \emph{Journal of Systems and Software}, vol. 123, pp. 176--189, 2017.

\bibitem{tldraw_2980}
{tldraw GitHub Repository}, ``{tldraw/tldraw Pull Request \#2495},'' \url{https://github.com/tldraw/tldraw/pull/2980}, 2024, accessed on March 20, 2024.

\bibitem{rombaut2023there}
B.~Rombaut, F.~R. Cogo, B.~Adams, and A.~E. Hassan, ``There's no such thing as a free lunch: Lessons learned from exploring the overhead introduced by the {Greenkeeper} dependency bot in npm,'' \emph{ACM Transactions on Software Engineering and Methodology}, vol.~32, no.~1, pp. 1--40, 2023.

\bibitem{mcintosh2011empirical}
S.~McIntosh, B.~Adams, T.~H. Nguyen, Y.~Kamei, and A.~E. Hassan, ``An empirical study of build maintenance effort,'' in \emph{Proceedings of the 33rd international conference on software engineering}, 2011, pp. 141--150.

\bibitem{jiang2015co}
Y.~Jiang and B.~Adams, ``Co-evolution of infrastructure and source code-an empirical study,'' in \emph{2015 IEEE/ACM 12th Working Conference on Mining Software Repositories}.\hskip 1em plus 0.5em minus 0.4em\relax IEEE, 2015, pp. 45--55.

\bibitem{wessel2023github}
M.~Wessel, J.~Vargovich, M.~A. Gerosa, and C.~Treude, ``{GitHub} actions: the impact on the pull request process,'' \emph{Empirical Software Engineering}, vol.~28, no.~6, p. 131, 2023.

\bibitem{valerp}
P.~Valenzuela-Toledo, A.~Bergel, T.~Kehrer, and O.~Nierstrasz, ``The hidden costs of automation: An empirical study on {GitHub} actions workflow maintenance replication package,'' \url{https://zenodo.org/record/12485092}, May 2024, version 1.0, Accessed: 2024-06-23.

\bibitem{nagappan2007using}
N.~Nagappan and T.~Ball, ``Using software dependencies and churn metrics to predict field failures: An empirical case study,'' in \emph{First International Symposium on Empirical Software Engineering and Measurement (ESEM 2007)}.\hskip 1em plus 0.5em minus 0.4em\relax IEEE, 2007, pp. 364--373.

\bibitem{shin2010evaluating}
Y.~Shin, A.~Meneely, L.~Williams, and J.~A. Osborne, ``Evaluating complexity, code churn, and developer activity metrics as indicators of software vulnerabilities,'' \emph{IEEE transactions on software engineering}, vol.~37, no.~6, pp. 772--787, 2010.

\bibitem{gall1998detection}
H.~Gall, K.~Hajek, and M.~Jazayeri, ``Detection of logical coupling based on product release history,'' in \emph{Proceedings. International Conference on Software Maintenance (Cat. No. 98CB36272)}.\hskip 1em plus 0.5em minus 0.4em\relax IEEE, 1998, pp. 190--198.

\bibitem{kalliamvakou2014promises}
E.~Kalliamvakou, G.~Gousios, K.~Blincoe, L.~Singer, D.~M. German, and D.~Damian, ``The promises and perils of mining {GitHub},'' in \emph{Proceedings of the 11th working conference on mining software repositories}, 2014, pp. 92--101.

\bibitem{Dabic:msr2021data}
O.~Dabic, E.~Aghajani, and G.~Bavota, ``Sampling projects in github for {MSR} studies,'' in \emph{18th {IEEE/ACM} International Conference on Mining Software Repositories, {MSR} 2021}.\hskip 1em plus 0.5em minus 0.4em\relax {IEEE}, 2021, pp. 560--564.

\bibitem{github_stars}
\BIBentryALTinterwordspacing
I.~GitHub, ``Saving repositories with stars,'' 2024, accessed: 2024-06-07. [Online]. Available: \url{https://docs.github.com/en/get-started/exploring-projects-on-github/saving-repositories-with-stars}
\BIBentrySTDinterwordspacing

\bibitem{githubblog2023}
GitHub, ``The state of open source and ai,'' \url{https://github.blog/2023-11-08-the-state-of-open-source-and-ai/}, 2023, [Online; accessed 7-May-2024].

\bibitem{weeraddana2023empirical}
N.~R. Weeraddana, X.~Xu, M.~Alfadel, S.~McIntosh, and M.~Nagappan, ``An empirical comparison of ethnic and gender diversity of {DevOps} and non-{DevOps} contributions to open-source projects,'' \emph{Empirical Software Engineering}, vol.~28, no.~6, p. 150, 2023.

\bibitem{vasilescu2015data}
B.~Vasilescu, A.~Serebrenik, and V.~Filkov, ``A data set for social diversity studies of {GitHub} teams,'' in \emph{2015 IEEE/ACM 12th working conference on mining software repositories}.\hskip 1em plus 0.5em minus 0.4em\relax IEEE, 2015, pp. 514--517.

\bibitem{nejati2023icse}
M.~Nejati, M.~Alfadel, and S.~McIntosh, ``Code review of build system specifications: Prevalence, purposes, patterns, and perceptions,'' in \emph{Proc. of the International Conference on Software Engineering (ICSE)}, 2023, p. 1213–1224.

\bibitem{agrawal1993mining}
R.~Agrawal, T.~Imieli{\'n}ski, and A.~Swami, ``Mining association rules between sets of items in large databases,'' in \emph{Proceedings of the 1993 ACM SIGMOD international conference on Management of data}, 1993, pp. 207--216.

\bibitem{scipy-kruskal}
S.~Developers, ``scipy.stats.kruskal,'' \url{https://docs.scipy.org/doc/scipy/reference/generated/scipy.stats.kruskal.html}, 2024, accessed: 2024-06-08.

\bibitem{charmaz2014grounded}
K.~Charmaz, ``Grounded theory in global perspective: Reviews by international researchers,'' \emph{Qualitative inquiry}, vol.~20, no.~9, pp. 1074--1084, 2014.

\bibitem{zimmermann2016card}
T.~Zimmermann, ``Card-sorting: From text to themes,'' in \emph{Perspectives on data science for software engineering}.\hskip 1em plus 0.5em minus 0.4em\relax Elsevier, 2016, pp. 137--141.

\bibitem{lo2023cognitive}
S.~B. Lo, A.~D. Svensson, C.~J. Presley, and B.~L. Andersen, ``A cognitive--behavioral model of dyspnea: qualitative interviews with individuals with advanced lung cancer,'' \emph{Palliative \& Supportive Care}, vol.~21, no.~6, pp. 1070--1077, 2023.

\bibitem{gallaba2022lessons}
K.~Gallaba, M.~Lamothe, and S.~McIntosh, ``Lessons from eight years of operational data from a continuous integration service: An exploratory case study of circleci,'' in \emph{Proceedings of the 44th international conference on software engineering}, 2022, pp. 1330--1342.

\bibitem{zhao2017impact}
Y.~Zhao, A.~Serebrenik, Y.~Zhou, V.~Filkov, and B.~Vasilescu, ``The impact of continuous integration on other software development practices: a large-scale empirical study,'' in \emph{2017 32nd IEEE/ACM International Conference on Automated Software Engineering (ASE)}.\hskip 1em plus 0.5em minus 0.4em\relax IEEE, 2017, pp. 60--71.

\bibitem{cassee2020silent}
N.~Cassee, B.~Vasilescu, and A.~Serebrenik, ``The silent helper: the impact of continuous integration on code reviews,'' in \emph{2020 IEEE 27th International Conference on Software Analysis, Evolution and Reengineering (SANER)}.\hskip 1em plus 0.5em minus 0.4em\relax IEEE, 2020, pp. 423--434.

\bibitem{rostami2023usage}
P.~Rostami~Mazrae, T.~Mens, M.~Golzadeh, and A.~Decan, ``On the usage, co-usage and migration of {CI/CD} tools: A qualitative analysis,'' \emph{Empirical Software Engineering}, vol.~28, no.~2, p.~52, 2023.

\bibitem{mazrae2023preliminary}
P.~R. Mazrae, A.~Decan, T.~Mens, and M.~Wessel, ``A preliminary study of github actions workflow changes,'' 2023.

\bibitem{delicheh2023preliminary}
H.~O. Delicheh, A.~Decan, and T.~Mens, ``A preliminary study of {GitHub} actions dependencies,'' in \emph{CEUR Workshop Proceedings}, vol. 3483, 2023, pp. 66--77.

\bibitem{decan2023outdatedness}
A.~Decan, T.~Mens, and H.~O. Delicheh, ``On the outdatedness of workflows in the {GitHub} actions ecosystem,'' \emph{Journal of Systems and Software}, vol. 206, p. 111827, 2023.

\bibitem{valenzuela2022evolution}
P.~Valenzuela-Toledo and A.~Bergel, ``Evolution of {GitHub} action workflows,'' in \emph{2022 IEEE International Conference on Software Analysis, Evolution and Reengineering (SANER)}.\hskip 1em plus 0.5em minus 0.4em\relax IEEE, 2022, pp. 123--127.

\bibitem{kinsman2021software}
T.~Kinsman, M.~Wessel, M.~A. Gerosa, and C.~Treude, ``How do software developers use {GitHub} actions to automate their workflows?'' in \emph{2021 IEEE/ACM 18th International Conference on Mining Software Repositories (MSR)}.\hskip 1em plus 0.5em minus 0.4em\relax IEEE, 2021, pp. 420--431.

\bibitem{saroar2023developers}
S.~G. Saroar and M.~Nayebi, ``Developers’ perception of github actions: A survey analysis,'' in \emph{Proceedings of the 27th International Conference on Evaluation and Assessment in Software Engineering}, 2023, pp. 121--130.

\bibitem{bouzenia2024resource}
I.~Bouzenia and M.~Pradel, ``Resource usage and optimization opportunities in workflows of {GitHub} actions,'' in \emph{Proceedings of the 46th IEEE/ACM International Conference on Software Engineering}, 2024, pp. 1--12.

\bibitem{koishybayev2022characterizing}
I.~Koishybayev, A.~Nahapetyan, R.~Zachariah, S.~Muralee, B.~Reaves, A.~Kapravelos, and A.~Machiry, ``Characterizing the security of {GitHub} {CI} workflows,'' in \emph{31st USENIX Security Symposium (USENIX Security 22)}, 2022, pp. 2747--2763.

\bibitem{valenzuela2023egad}
P.~Valenzuela-Toledo, A.~Bergel, T.~Kehrer, and O.~Nierstrasz, ``{EGAD}: A moldable tool for {GitHub} action analysis,'' in \emph{2023 IEEE/ACM 20th International Conference on Mining Software Repositories (MSR)}.\hskip 1em plus 0.5em minus 0.4em\relax IEEE, 2023, pp. 260--264.

\bibitem{DBLP:conf/iwst/Valenzuela-Toledo23}
P.~Valenzuela{-}Toledo, A.~Bergel, T.~Kehrer, and O.~Nierstrasz, ``Exploring {GitHub} actions through {EGAD:} an experience report,'' in \emph{Proceedings of the International Workshop on Smalltalk Technologies, Lyon, France; August 29th-31st, 2023}, ser. {CEUR} Workshop Proceedings, vol. 3627.\hskip 1em plus 0.5em minus 0.4em\relax CEUR-WS.org, 2023.

\bibitem{cardoen2024dataset}
G.~Cardoen, T.~Mens, and A.~Decan, ``A dataset of {GitHub} actions workflow histories,'' in \emph{2024 IEEE/ACM 21st International Conference on Mining Software Repositories (MSR)}.\hskip 1em plus 0.5em minus 0.4em\relax IEEE, 2024, pp. 677--681.

\end{thebibliography}
\bibliographystyle{IEEEtran}

\end{document}